\documentclass[preprint,12pt]{elsarticle}

\usepackage{amssymb}
\usepackage{float}
\usepackage{graphicx}
\usepackage[utf8]{inputenc}
\usepackage{subcaption}
\usepackage{booktabs}
\usepackage{multimedia}
\usepackage[english]{babel}
\usepackage[T1]{fontenc}
\usepackage{subcaption}
\usepackage{url}

\journal{Physica A}

\begin{document}

\begin{frontmatter}

\title{Congestion fronts of diffusing particles}

\author[add1]{B. Burger}
\author[add1,add2]{H. J. Herrmann}

\address[add1]{IfB, HIT G23.1, ETH Zürich, 8093 Zürich, Switzerland}
\address[add2]{Departamento de Física, Universidade Federal do Ceará, 60451-970 Fortaleza, Ceará, Brazil}

\newpage

\begin{abstract}
  We study two new models of two particle species invading a surface from opposite sides. Collisions of particles of different species lead to the formation of \textit{congestion fronts}. One of the models implements a reversible process whereas in the other model the congestion front forms irreversibly. For both models we find that the congestion fronts are self-affine but with different roughness exponents. For low densities the system does not congest and we find a phase transition between a phase of freely moving particles and a congestion phase.
\end{abstract}

\end{frontmatter}

\section*{Keywords}
Diffusion; Growth Models; Roughness

\section{Introduction}
Many forms of reaction-diffusion type processes have been studied numerically in the past \cite{turk1991generating,hattne2005stochastic,grindrod1996theory,britton1986reaction}. One of the most common settings is the following: One starts with two spatially separated reagents A and B that move diffusively. If the reagents coexist at the same place they react and produce a product C. The region where the production of C is nonzero is called a \textit{reaction front} \cite{galfi1988properties}. As such, reaction-diffusion processes have found a wide range of applications in biology \cite{britton1986reaction,kondo2010reaction}, physics \cite{colizza2007reaction} and ecology \cite{cantrell2004spatial}.

We consider here the situation in which one species is continuously injected from the right and the other from the left. A similar situation is encountered when pedestrians move in opposite directions against each other in a corridor \cite{helbing2001self,helbing2005self}, although that instead of bumping into each other they rather tend to form lanes \cite{helbing2005self,helbing1995social}.

Here we introduce a model of two randomly advancing particle species that collide with each other, eventually leading to the congestion of the system and with it the formation of an interface. The interface is smoothened in the version of our model that incorporates surface relaxation due to lateral movement of the particles, similar to random deposition with surface relaxation (RDSR) \cite{family1986scaling,mattos2006new}. We simulate different system sizes, find the underlying formation processes and test the resulting interfaces for fractality and self-affinity.

\section{Method}
Suppose that there are two groups of drunken people at a music festival. After the concert, the audience from stage A wants to switch to stage B, whereas stage B's audience switches places to stage A. The drunkards approach each other and as the audiences start to collide, the system either congests, or everyone arrives at their desired stage. Mimicking such a scenario, we simulate two particle species invading a two-dimensional surface from two opposite sides. We consider a square lattice where one species is injected from the right and moves to the left and the other is injected from the left and moves to the right. Each species is labeled with a colour, i.e. blue and red and has the option to move forward, upward or downward, as it is not allowed to move backward to impose a particle drift. We denote the probabilities to move in the respective direction with $p_{advance}$, $p_{up}$ and $p_{down}$. To avoid asymmetries, we furthermore impose the restriction
\begin{equation}
  p_{up} = p_{down}.
  \label{eq:symmetry}
\end{equation}
In the transverse direction we apply periodic boundary conditions and in the longitudinal direction particles disappear as soon as they arrive at the other side of the system.

The simulation operates in cycles to simulate the particle dynamics. Each cycle starts with a release of particles characterized by the release rate $p_{release}$. The release rate defines the probability with which a particle is released into the system on each site of the left (blue) and the right (red) edge. After releasing particles into the system, we need to move the particles. For a system of $N$ particles, we choose $N$ particles at random, some of them eventually multiple times, and move them in one of the allowed directions with the respective probability. If a particle tries to move to an already occupied site, it avoids a collision by performing no movement instead. Each time a particle is selected for an update, it can just make one attempt to move, leading to the condition
\begin{equation}
 p_{advance} + p_{up} + p_{down} = 1.
 \label{eq:unity}
\end{equation}

\begin{figure}
  \centering
  \begin{subfigure}{.5\textwidth}
    \centering
    \includegraphics[width = .9\linewidth]{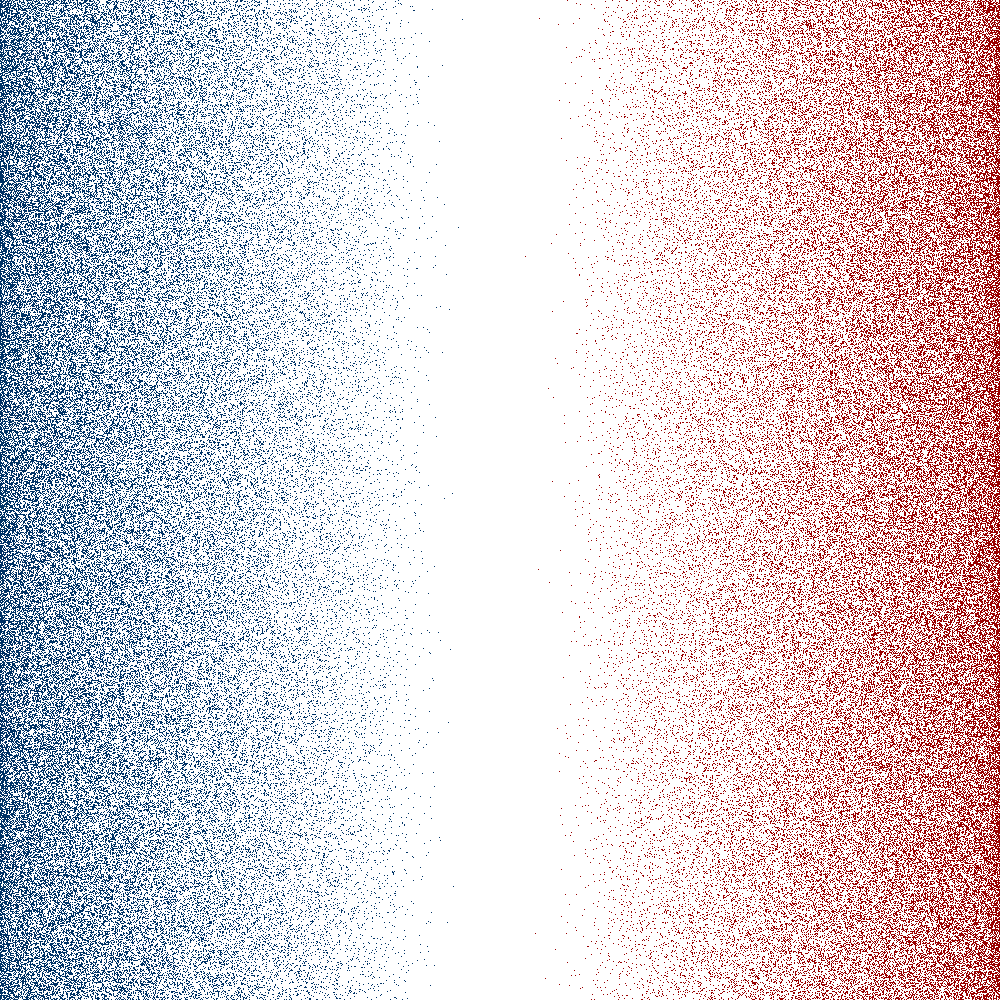}
    \caption{$t=500$}
  \end{subfigure}%
  \begin{subfigure}{.5\textwidth}
    \centering
    \includegraphics[width = .9\linewidth]{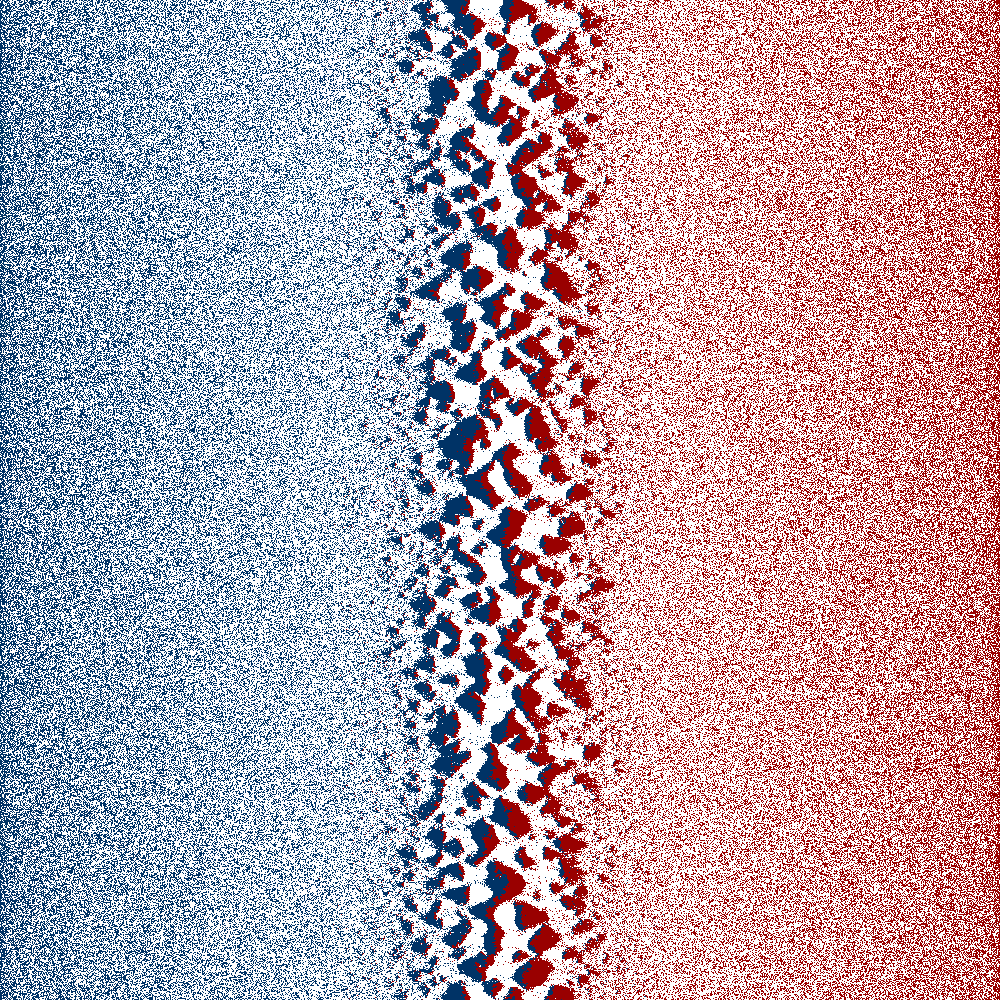}
    \caption{$t=1'000$}
  \end{subfigure}
  \begin{subfigure}{.5\textwidth}
    \centering
    \includegraphics[width = .9\linewidth]{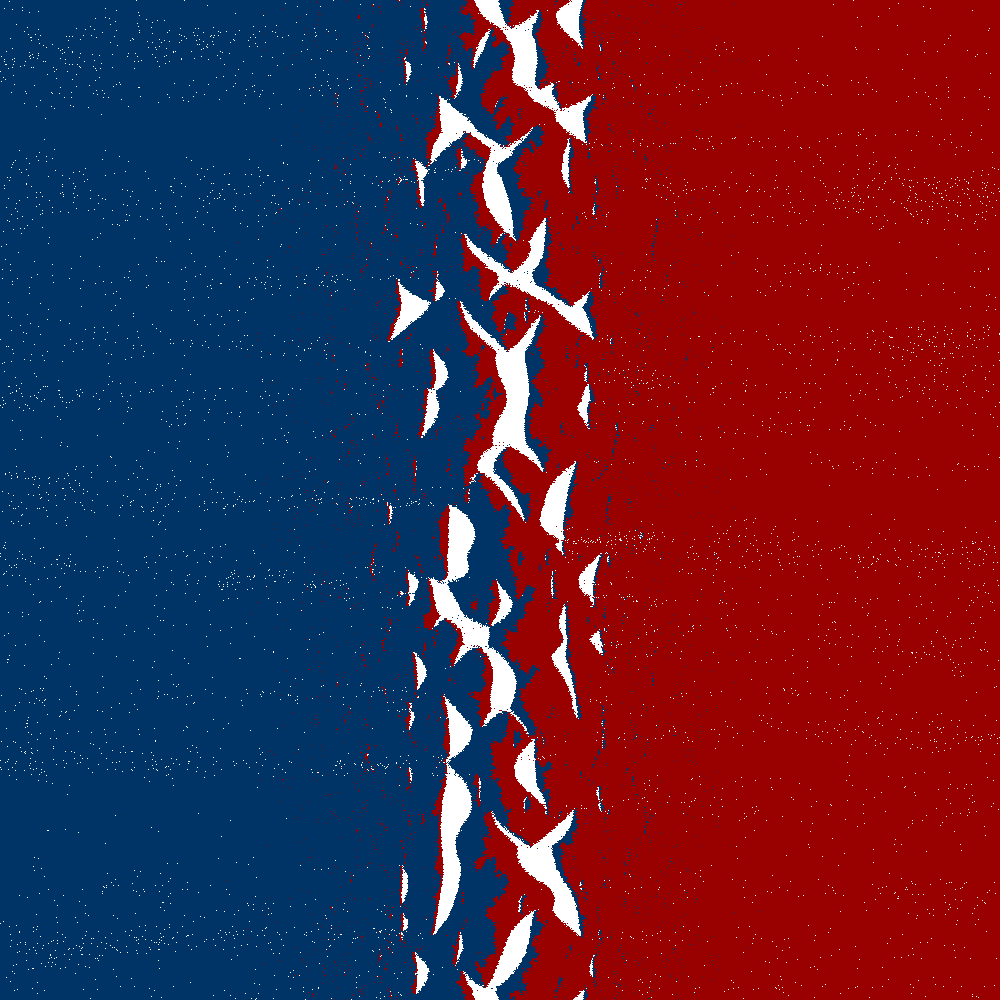}
    \caption{$t=2'000$}
  \end{subfigure}%
  \begin{subfigure}{.5\textwidth}
    \centering
    \includegraphics[width = .9\linewidth]{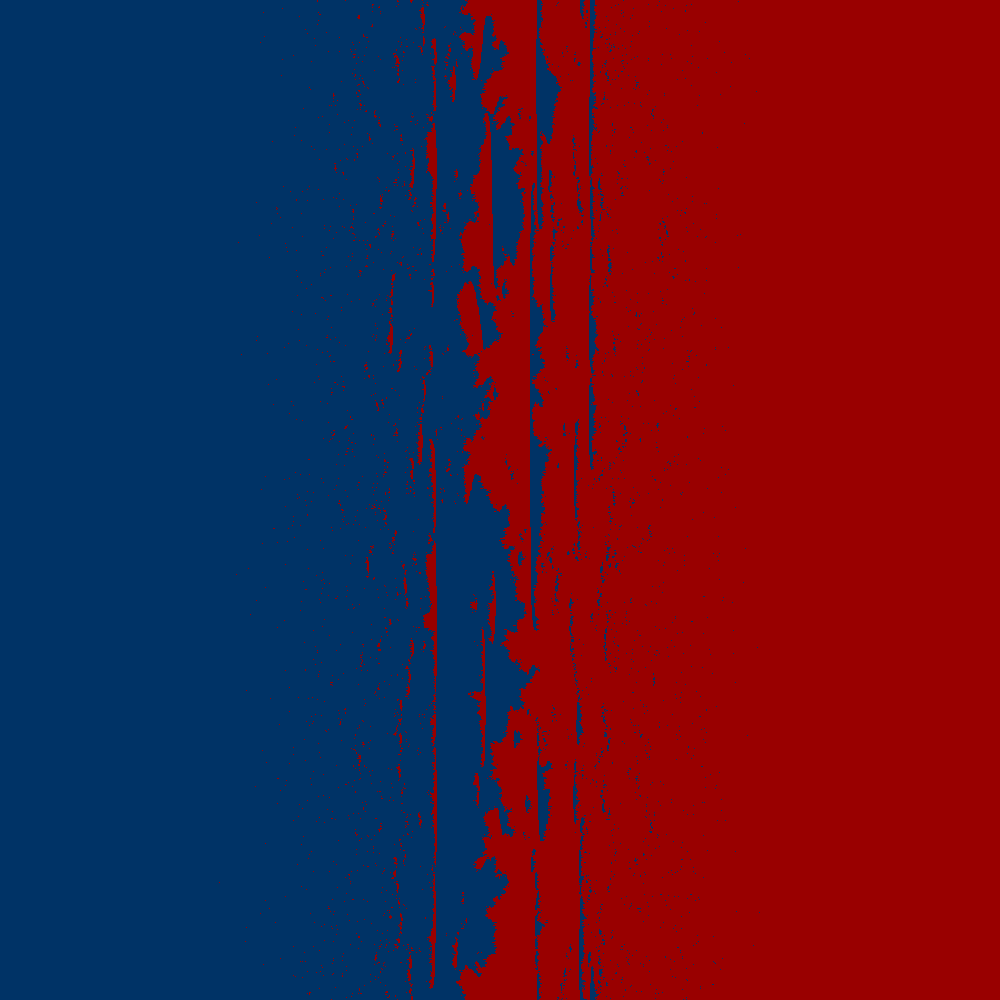}
    \caption{$t=1'000'000$}
  \end{subfigure}
  \caption{System at different times for $p_{advance}=0.8$ and $r_{release}=0.8$ after $t$ cycles. The white regions are empty.}
  \label{fig:formation}
\end{figure}

Collisions of particles of different species results in the formation of clusters of immobile particles. The \textit{congestion fronts} are the lines through these clusters separating the two colours. These clusters usually grow, morph and fuse with other clusters until the whole system congests, yielding an interface between red and blue that spans across the system. We call this spanning interface the \textit{immutable} congestion front (Fig.~\ref{fig:formation}), because once it has formed, it will not change anymore.

\begin{figure}
  \centering
  \begin{subfigure}[t]{\textwidth}
    \includegraphics[width = .9\linewidth]{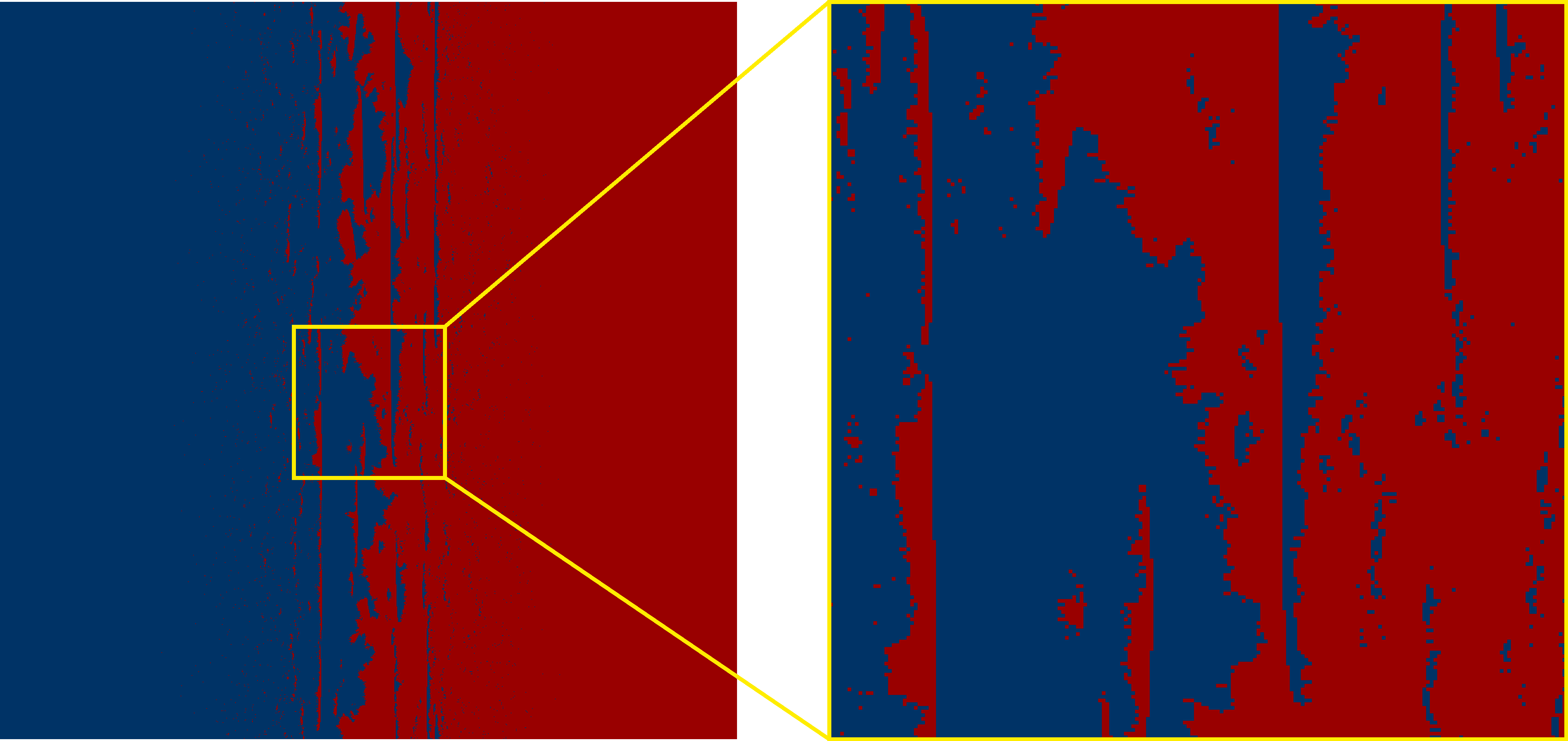}
    \captionsetup{width=0.8\linewidth}
    \caption{Immutable congestion front.}
  \end{subfigure}
  \begin{subfigure}[t]{\textwidth}
    \includegraphics[width = .9\linewidth]{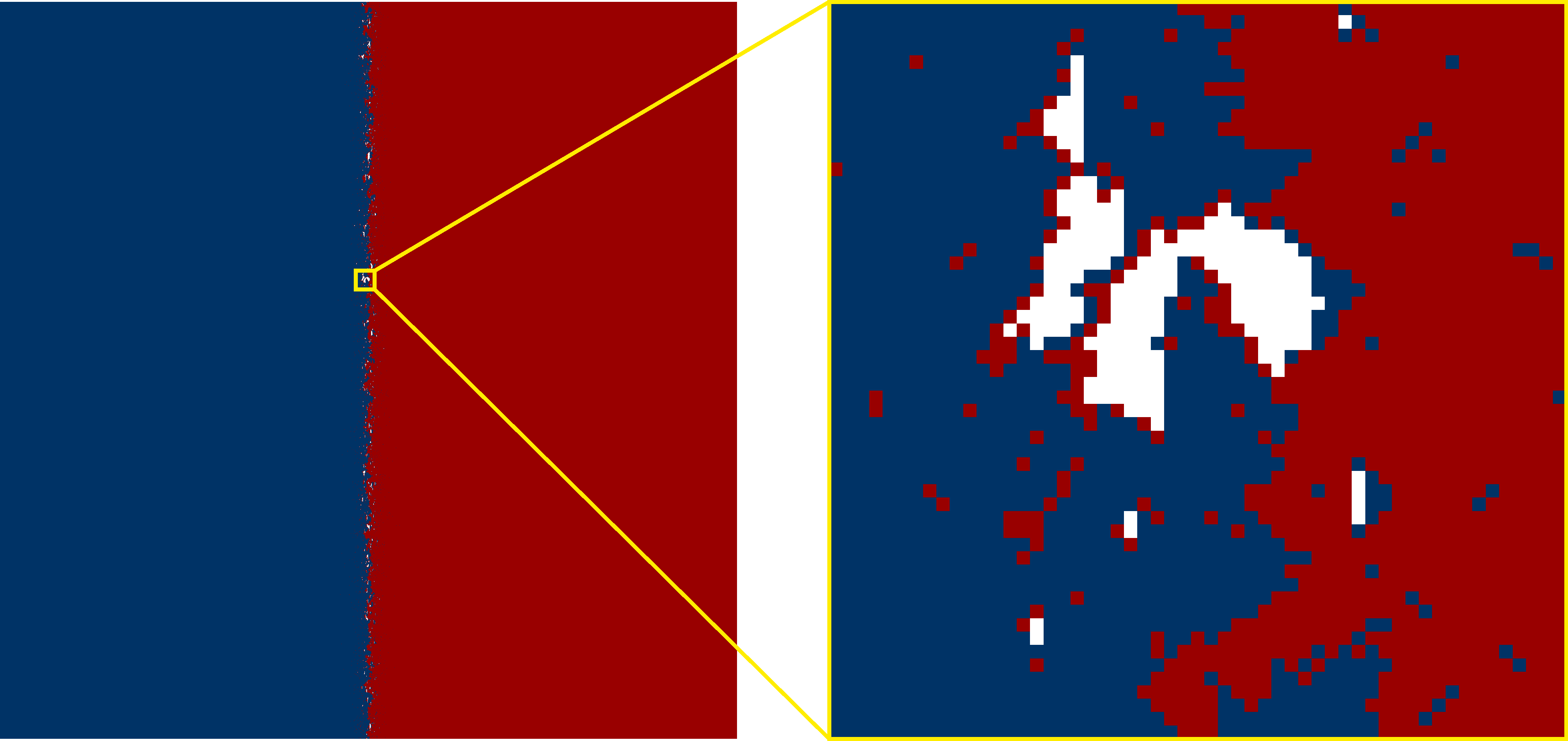}
    \captionsetup{width=0.8\linewidth}
    \caption{Irreversible congestion front.}
  \end{subfigure}
  \caption{Congestion fronts for $p_{advance} = p_{release} = 0.8$. The right side is a zoom of a small part of the congestion front shown at the left.}
  \label{fig:congestion_front}
\end{figure}

The model we have specified so far allows for a reversible congestion front formation. Another physical model, similar to reaction-diffusion type processes, is when two particles of different species coming in contact react, or stick together, and the product immediately deposits on the surface. To explore this setting, as well as to better understand the effect of restructuring on the final congestion front, we introduce an irreversible model. We require that as soon as two differently coloured particles collide, they stop and will never move again. This guarantees the irreversibility of the congestion front and completely suppresses surface relaxation (Fig.~\ref{fig:congestion_front}). To distinguish the congestion front of the reversible model from the congestion front of the irreversible model we call the spanning congestion front of the reversible model the \textit{immutable congestion front} and the congestion front of the irreversible model the \textit{irreversible congestion front}.

The macroscopic outcome of the simulation is determined by the release rate $p_{release}$ and the forward movement probability $p_{advance}$, which in turn implies the values of $p_{up}$ and $p_{down}$ by using Eqs. \ref{eq:symmetry} and \ref{eq:unity}. Each system is thus specified through $p_{release}$ and $p_{advance}$.

\section{Results}
\subsection{Spatio-temporal evolution}
Before the quantitative analysis of the congestion front, we first qualitatively discuss the patterns involved in the congestion front formation process (Fig.~\ref{fig:formation}).

\subsubsection{Reversible model}
 The cluster dynamics revolves around the two-colour interfaces that form if particles collide. Particles are able to leave a cluster by moving laterally, leading to streams of particles emerging at the two edges of the congestion front of a cluster. These streams can contribute to the mass of another cluster, where they might smoothen the interface due to lateral movement. They can also escape this cluster and move on to the next one, until the system congests and the particles can not escape anymore. The interplay between a cluster's continuous mass loss, particle accumulation and cluster reconfiguration determines the outcome of the simulation. If cluster growth and cluster fusion is faster than the mass loss, the system finally clogs and a spanning congestion front forms.

The particle accumulation rate of a cluster depends on its cross section in the transverse direction. A larger cross section offers a larger area that is able to catch incoming particles. On the other hand, the influence of the cross section on the streams of leaving particles is small. A cluster with a cross section larger than a critical value is accumulating particles faster than losing them, resulting in a positive feedback loop and finally the congestion of the system.

\begin{figure}
  \centering
  \includegraphics[width=\textwidth]{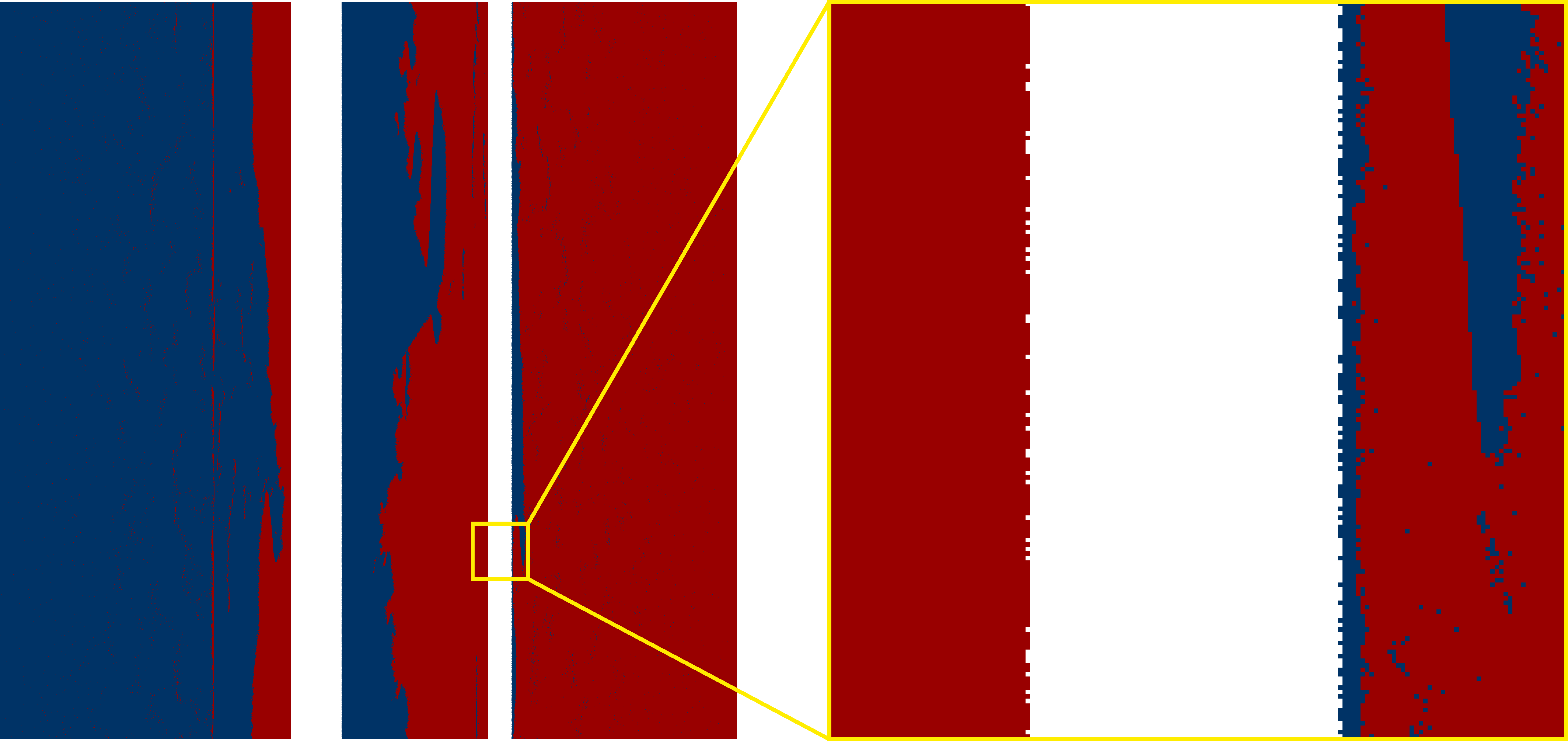}
  \caption{System of multiple congestion fronts for $p_{release} = p_{advance} = 0.4$. The white areas are empty spaces that will never fill up because each immutable congestion front spans the whole system and is therefore an insurmountable obstacle. For $t \rightarrow \infty$, the interfaces between white and coloured areas are straight lines on which there is one layer of diffusing particles.}
  \label{fig:white_particles}
\end{figure}

If we lower the density and increase the probability for lateral movement, we observe the formation of multiple immutable congestion fronts, separated by areas of empty space (Fig.~\ref{fig:white_particles}). Each spanning congestion front is an insurmountable obstacle, cutting off the particle supply needed to fill up empty areas. The interface between an empty and a coloured area becomes a straight line because the particles will fill up all local minima after sufficient time. It is however only straight to the eye: On a microscopic level, the last layer of particles is only partially filled (Fig.~\ref{fig:white_particles} (b)). We call particles in a line that will never completely fill up \textit{spare particles}. The line in front of spare particles is always completely occupied and hinders them to advance, but the particles will always have the option to move laterally. For this reason the spare particles will never settle down and there will always be some movement of spare particles in the system.

\begin{figure}
  \centering
  \includegraphics[width=.9\textwidth]{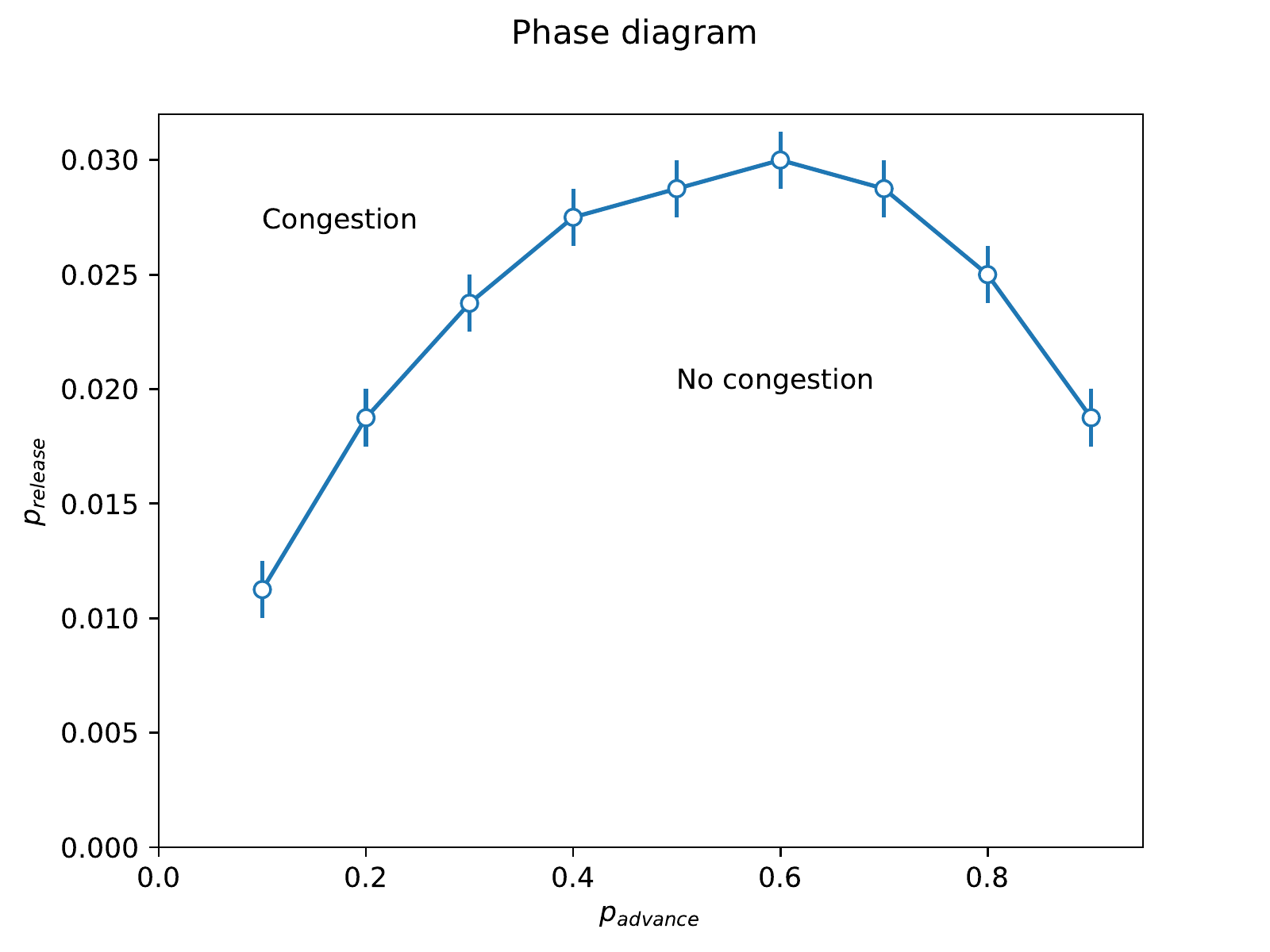}
  \caption{Phase diagram for the transition between the clogging and the non-clogging phase.}
  \label{fig:phase_diagram}
\end{figure}

If we lower the density even more, configurations show no signs of clogging anymore. Intuitively it is clear that for very low densities the particles will always find a way to evade each other, which indicates that there must exist a phase where the system does not clog. The existence of two phases suggests the existence of a phase transition. We went up to $5 \times 10^6$ cycles to check if a given configuration clogs: If it does not, we classify the configuration as belonging to the non-congestion phase. With this methodology, we establish the phase diagram between the clogging and the non-clogging phase shown in Fig. \ref{fig:phase_diagram}.

\subsubsection{Irreversible algorithm}
For the irreversible algorithm other forms of areas remain empty. Small areas of empty space (Fig.~\ref{fig:congestion_front} (d)) survive when they are surrounded by already collided particles that will never move anymore and by particles that are not allowed to move backward as seen in Fig.~\ref{fig:congestion_front} (d).

\subsubsection{Ballistic limit}
In the limit $p_{advance} = 1$, the particles must always move forward, unable to change their horizontal trajectory. For this reason in this limit both the reversible and the irreversible model are identical. This case resembles the rain model \cite{kertesz1994self}, but is not completely identical because we are not looking at surface growth in the classical sense.

\subsection{Quantitative analysis}
The quantitative analysis of the immutable congestion front comprises testing the interface for fractality and roughness. We test for fractality by applying a yardstick measurement \cite{tricot1988evaluation} and we also plot the interface length $M$ against the system size $L$, where we would expect a power law $M \propto L^{d_f}$ in the case of fractality. The yardstick measurements as well as the $M$-$L$ relation exhibit strong curvature in a log-log plot over three orders of magnitude, with maximum system length $L = 4000$, for different parameters and both the reversible and the irreversible model. This leads to the conclusion that the congestion front is not fractal for any of the two models.

The question remains whether the congestion front is self-affine for any of the models. We will consider two different interfaces: the \textit{longest spanning congestion front} and the set of all interface pieces, which we call the \textit{total congestion front}. The interface width \begin{equation}W = \sqrt{\frac{1}{N}\sum_i(\bar{x}-x_i)^2},\label{eq:width}\end{equation} where $\bar{x} = \sum_ix_i$ denotes the mean position of the interface, depends on the system's length L as a power law \begin{equation}W \propto L^\alpha\label{eq:width_length_relation}\end{equation} if the interface is self-affine, $\alpha$ being the roughness exponent \cite{kertesz1994self}. For each set of parameters we measured the interface width averaged over at least $300$ systems for each size. We then plotted the mean width double-logarithmically in dependency of the system size (Fig.~\ref{fig:measurements}). To determine the roughness exponent $\alpha$ we fitted the data to Eq.~\ref{eq:width_length_relation}. We calculated the error on the roughness exponent by evaluating the lines with maximum and minimum slopes that were still fitting the data within their individual statistical error bars. We will not consider $p_{release} < 0.2$, because in or close to the non-congestion phase shown in Fig.~\ref{fig:phase_diagram}, congestion fronts do not appear.

\subsubsection{Ballistic limit}
In the limit $p_{advance} = 1.0$, both the reversible and irreversible model as well as the longest spanning and total congestion fronts are identical. Therefore every system crosses over to the ballistic limit for $p_{advance} = 1.0$. The measurement of the interface width exhibits a convincing power law with a roughness exponent $\alpha = 0.55 \pm 0.04$. This roughness exponent seems to belong to the universality class of KPZ \cite{kardar1986dynamic} and ballistic deposition \cite{meakin1986ballistic}.

\subsubsection{Longest spanning, irreversible congestion front}
For the irreversible model the interface width measurement clearly yields a power law for parameters $0.2 \leq p_{advance} = p_{release} \leq 0.8$, which indicates that the longest spanning immutable congestion front is self-affine with a roughness exponent $\alpha = 0.31 \pm 0.01$ (Fig.~\ref{fig:measurements} (a)). The roughness exponent is remarkably small. We are not aware of any other interface growth model yielding an exponent consistent with such a roughness exponent.

\subsubsection{Total, irreversible congestion front}
As seen in Fig.~\ref{fig:measurements}~(a), the total, irreversible congestion front shows no power law over the whole range. However, for $L \geq 800$ we can see that there might be an asymptotic power law, but the measurement range spans too few order of magnitudes to give a conclusive answer. If we assume that the power law holds and evaluate the roughness exponent, we get $\alpha = 0.62 \pm 0.05$ for $0.2 \leq p_{advance} = p_{release} \leq 0.8$. Interestingly, the roughness exponent is consistent with the exponent of directed percolation depinning (DPD): In DPD, quenched disorder is represented as a fraction $p$ of blocked cells on a cubic lattice. The interface is allowed to move freely on unblocked cells, whereas blocked cells hinder the interface to propagate. For $p=p_c$ the interface is pinned and its roughness exponent is $\alpha=0.633 \pm 0.001$ \cite{barabasi1995fractal}. This setting is similar to the irreversible model. As soon as two particles collide, their interface behaves like a blocked cell. It hinders the propagation and it is quenched, as the collided particles are not allowed to ever move again. The key difference is the formation mechanism. In DPD we have a percolation model with interface growth, whereas the irreversible congestion front model consists of particles that move randomly, leading to interface formation.

\subsubsection{Longest spanning, immutable congestion front}
For the reversible model, the width measurement of the longest spanning congestion front shows self-affine behaviour (Fig.~\ref{fig:measurements} (b)) for parameters $0.2 \leq p_{advance} = p_{release} \leq 0.8$. These configurations yield a roughness exponent $\alpha = 0.77 \pm 0.03$. This exponent is consistent with the exponent of the moving interface, where $p<p_c$, in DPD \cite{barabasi1995fractal}. In this regime the interface is not pinned and propagates within the whole system. Several numerical, experimental and theoretical studies have explored models of surface roughening in disordered media \cite{kessler1991interface,csahok1993dynamics}, fluid displacement in disordered media \cite{rubio1989self} or the kinetic interfaces of patchy particles \cite{araujo2015kinetic} and obtained roughness exponents $0.71 \leq \alpha \leq 0.75$, which is in the directed percolation universality class.

\subsubsection{Total, immutable congestion front}
\begin{figure}
  \centering
  \begin{subfigure}[t]{.8\textwidth}
      \includegraphics[width = \linewidth]{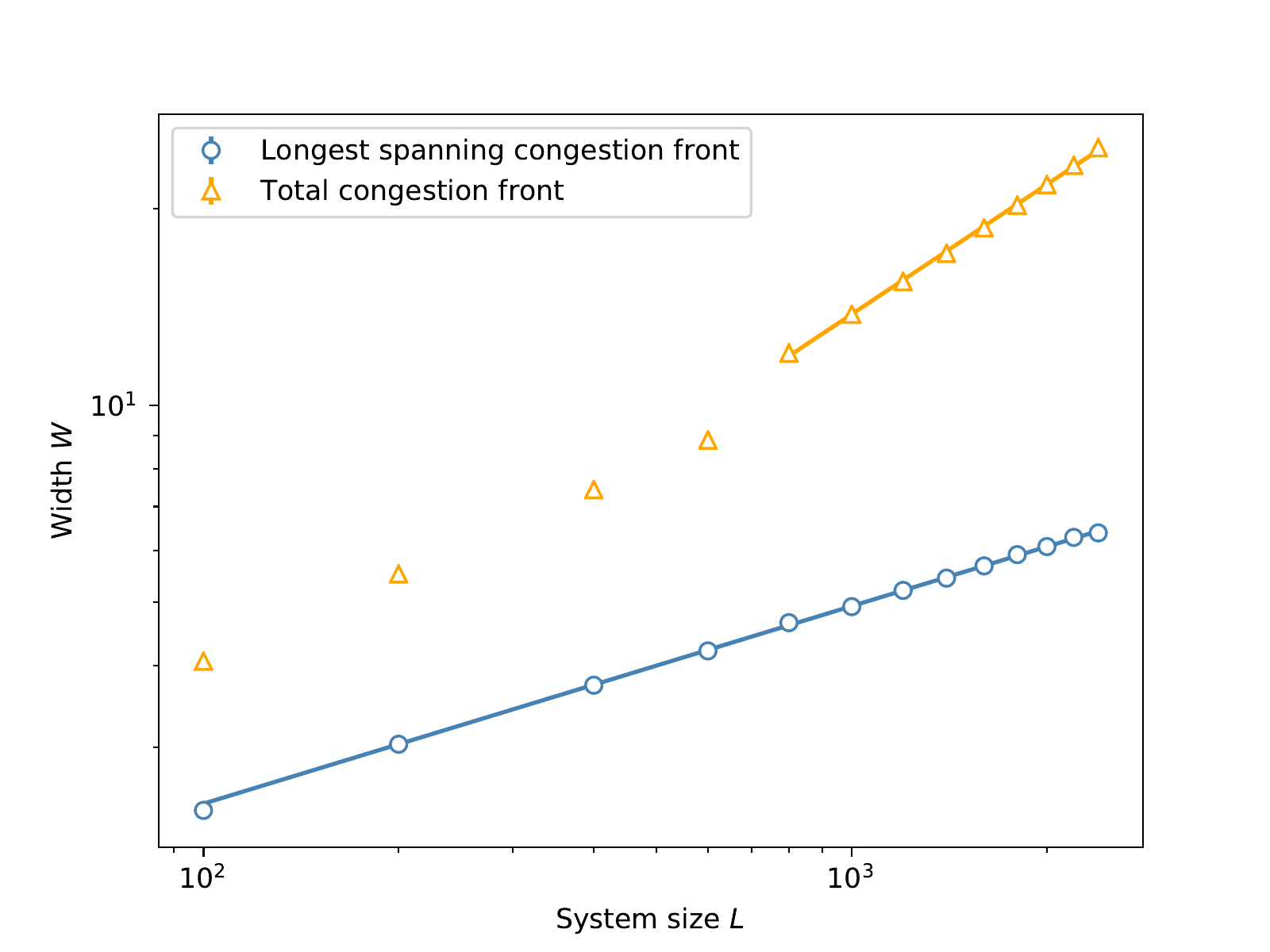}
      \caption{Irreversible model, $p_{release}=0.4$, $p_{advance}=0.4$.}
  \end{subfigure}
  \begin{subfigure}[t]{.8\textwidth}
  \includegraphics[width = \linewidth]{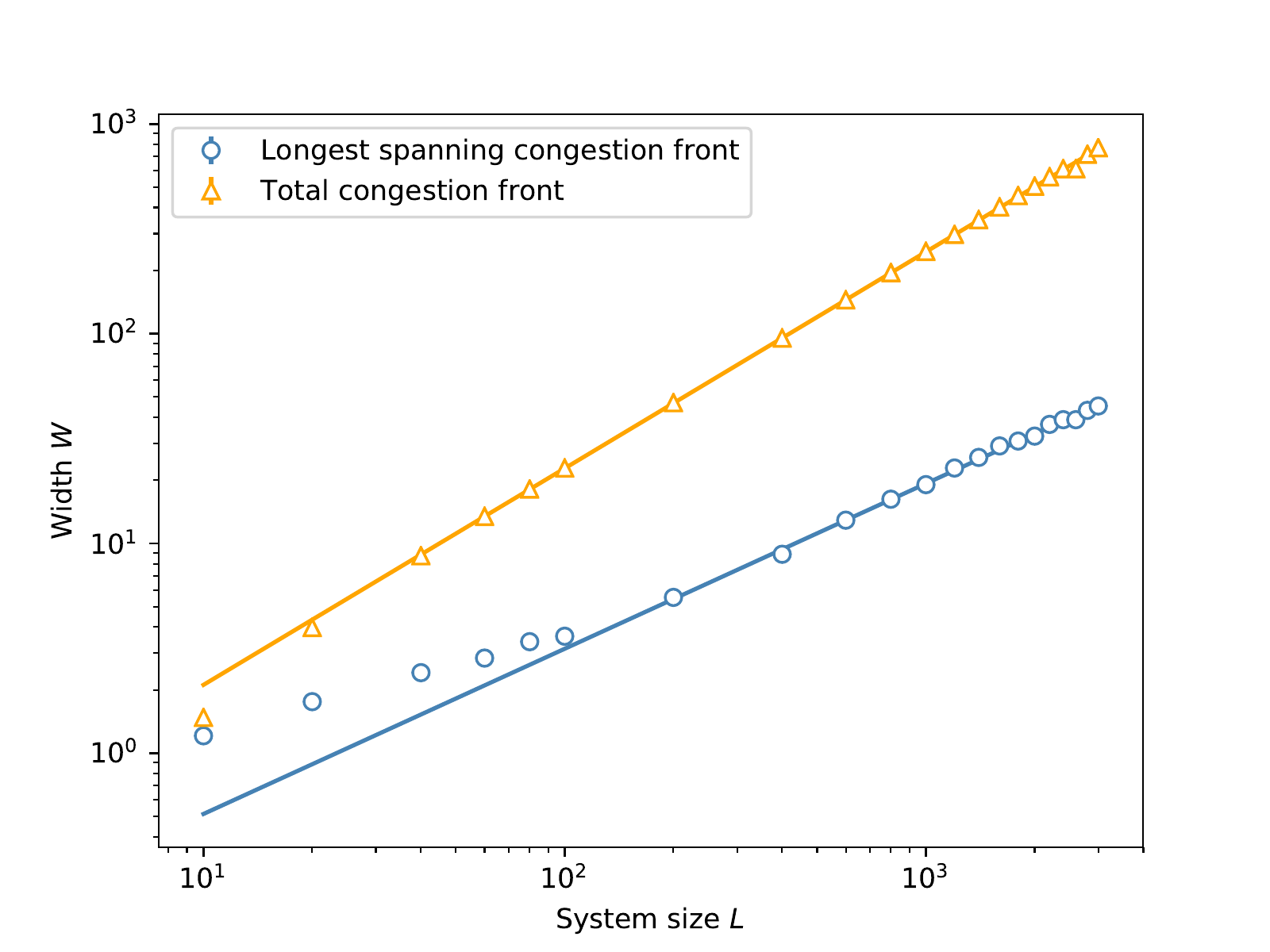}
  \caption{Reversible model, $p_{release}=0.4$, $p_{advance}=0.4$.}
\end{subfigure}
\caption{Examples of width measurements for both models.}
\label{fig:measurements}
\end{figure}

We observe power law behaviour and obtain $\alpha = 1.05 \pm 0.02$ for $0.2 \leq p_{advance} = p_{release} \leq 0.8$ (Fig.~\ref{fig:measurements} (b)). Note that the roughness exponent $\alpha \approx 1$ might imply fractality, but the box-counting method we applied to test this did not show convincing scaling behaviour. Therefore they have a finite density and are rather two-dimensional instead of fractal.

\begin{figure}
  \centering
  \includegraphics[width=.9\textwidth]{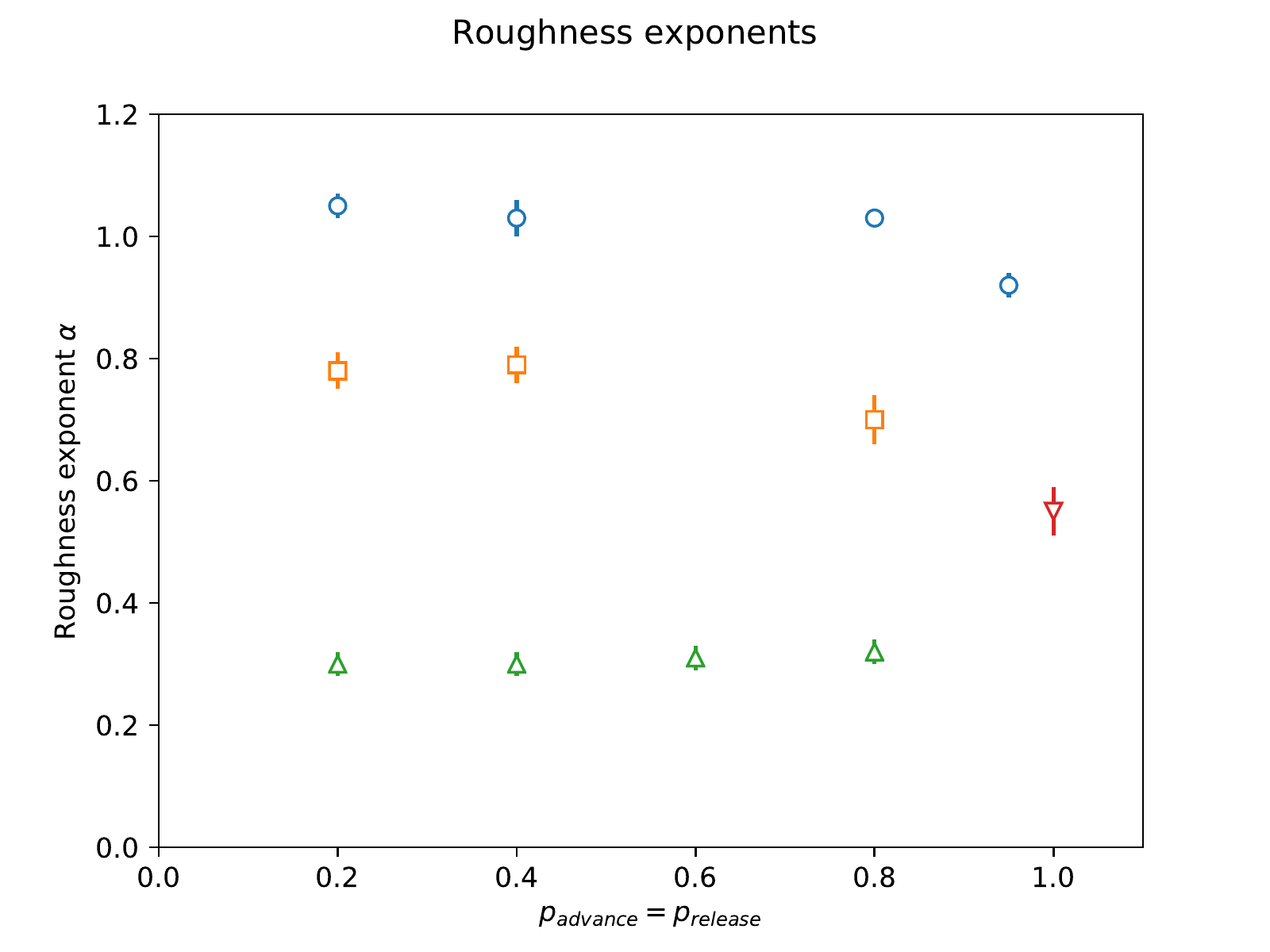}
  \caption{Measured roughness exponents $\alpha$ for the total immutable congestion front (blue circles), the longest spanning immutable congestion front (orange squares), the longest spanning irreversible collision front (green triangles) and the ballistic limit $p_{advance} = 1.0$ (red inverted triangle).}
  \label{fig:roughness_results}
\end{figure}

We summarize the results of the roughness measurements in Fig.~\ref{fig:roughness_results}, where we can see the dependency of the roughness exponent $\alpha$ on the model parameters. We choose $p_{advance} = p_{release}$ for the simplicity of presenting a two-dimensional plot.

\section{Discussion}
We defined a new congestion model through diffusing particles with drift. A non-congestion phase and a congestion phase are distinguished in a phase diagram. We find different roughness exponents, depending on whether the congestion front forms reversibly or not. The roughness exponent for the immutable congestion front is found to be $0.77 \pm 0.03$ and thus in the universality class of directed percolation depinning \cite{barabasi1995fractal,kessler1991interface,csahok1993dynamics,rubio1989self}. The roughness exponent $\alpha = 0.31 \pm 0.01$ of the longest spanning, irreversible congestion front is remarkably small and we know no related universality class in two dimensions with the same roughness exponent. For the total immutable congestion front we find $\alpha = 1.05 \pm 0.02$, the interface thus being dense. In the limit of no lateral movement, where $p_{advance} = 1.0$, all models are identical with roughness exponent $\alpha = 0.55 \pm 0.04$, being in the universality class of KPZ \cite{kardar1986dynamic} and ballistic deposition \cite{meakin1986ballistic}.

\end{document}